\documentclass[12pt,preprint]{aastex}

\usepackage{epsfig}
\usepackage{psfig}
\usepackage{lscape}

\slugcomment{}

\shorttitle{}
\shortauthors{Hui et al.}

\begin{document}

\title{Differences between radio-loud and radio-quiet $\gamma-$ray pulsars as revealed by \emph{Fermi}} 

\author{
C.~Y.~Hui\altaffilmark{1},
Jongsu Lee\altaffilmark{1},
J.~Takata\altaffilmark{2},
C.W. Ng\altaffilmark{3}
K.~S.~Cheng\altaffilmark{3},
}
\email{cyhui@cnu.ac.kr}
\altaffiltext{1}{Department of Astronomy and Space Science, Chungnam National University, Daejeon, Republic of Korea}
\altaffiltext{2}{ Institute of Particle physics and Astronomy, Huazhong University of Science and Technology}
\altaffiltext{3}{Department of Physics, University of Hong Kong, Pokfulam Road, Hong Kong}


\begin{abstract}
By comparing the properties of non-recycled radio-loud $\gamma-$ray pulsars and radio-quiet $\gamma-$ray pulsars, 
we have searched for the differences between these two populations. 
We found that the $\gamma-$ray spectral curvature of radio-quiet pulsars can be larger than that of radio-loud pulsars.
Based on the full sample of non-recycled $\gamma-$ray pulsars, their distributions of the magnetic field strength at the light cylinder 
are also found to be different. We notice that this might be resulted from the observational bias. 
In re-examining the previously reported difference of $\gamma-$ray-to-X-ray flux ratios,  
we found the significance can be hampered by their statistical uncertainties. 
In the context of outer gap model, we discuss the expected properties of these two populations and compare with the possible differences identified 
in our analysis. 
\end{abstract}

\keywords{gamma rays: stars --- Pulsars: general}

\section{Introduction}
Before the launch of \emph{Fermi} $\gamma-$ray Space Telescope, our understanding of 
$\gamma-$ray pulsars was very limited. Its predecessor \emph{Compton} Gamma-Ray 
Observatory (CGRO) has only detected seven pulsars in MeV-GeV regime throughout 
its almost nine years life-time (Thompson 2008). Among them, there is a special member, 
Geminga (PSR~J0633+1746), which was the only known radio-quiet $\gamma-$ray pulsar in the 
pre-\emph{Fermi} era (see Bignami \& Caraveo 1996 for a review). 

With its much improved sensitivity and accurate source localization, the Large Area 
Telescope (LAT) onboard \emph{Fermi} has expanded the $\gamma-$ray pulsar population 
considerably shortly after its operation (Abdo et al. 2009a,b). 16 new $\gamma-$ray
pulsars have been discovered through blind searches with just $\sim4.5$ month data 
(Abdo et al. 2009a). Currently, there are 205 $\gamma-$ray pulsars have been 
detected by LAT.\footnote{For updated statistics, please refer to 
\scriptsize{https://confluence.slac.stanford.edu/display/GLAMCOG/Public+List+of+LAT-Detected+Gamma-Ray+Pulsars}.}

In the second \emph{Fermi} LAT pulsar catalog (2PC Abdo et al. 2013), the detailed properties of 
117 pulsars detected at energies $>100$~MeV with three years data are reported. 
It comprises 42 radio-loud pulsars, 35 radio-quiet pulsars and 40 millisecond pulsars 
(Abdo et al. 2013)\footnote{Radio-loud or radio-quiet in 2PC is defined by 
whether its radio flux density at 1.4~GHz is larger or smaller than 30~$\mu$Jy}. 

Establishing radio-quiet $\gamma-$ray pulsars as a definite class is one of the triumphs 
of \emph{Fermi}. Different from the radio-loud cases, they can only be detected through 
blind pulsation searches at high energies. Apart from the high sensitivity of LAT, the expansion 
of radio-quiet pulsar population also thanks to the improvement of searching techniques
(e.g. Kerr 2011). 

About 30$\%$ of the known $\gamma-$ray pulsars are radio-quiet. Taking the selection effects  
into account, this fraction can be even larger. Sokolova \& Rubtsov (2016) have estimated that 
the intrinsic fraction of radio-quiet $\gamma-$ray pulsars can be as large as $\sim70\%$. 
Such large fraction of radio-quietness 
imposes strict constraints on the geometry and mechanism of the pulsar emission. This implies 
the $\gamma-$rays are originated from the outer magnetosphere and form a fan beam 
(see Cheng \& Zhang 1998; Takata et al. 2006, 2008). In comparison with the narrow cone-like 
radio beam originated from the polar cap region, this makes the detection of $\gamma-$ray 
pulsation less sensitive to the emission and viewing geometry. 

As the sample sizes of radio-quiet $\gamma-$ray pulsars and the non-recycled radio-loud $\gamma-$ray 
pulsars are now comparable, a deeper insight of their nature can be gained by comparing their physical and emission 
properties. Marelli et al. (2011,2015) and Marelli (2012) have shown that the $\gamma-$ray-to-X-ray flux ratios 
of radio-quiet population are higher than that of radio-loud ones. While these works did not found  
any solid evidence for the difference between these two populations neither in terms of the 
physical properties (e.g. magnetic field) nor in $\gamma-$ray regime, 
Marelli et al. (2015) suggest this implies the X-ray emission of the radio-quiet population is 
generally fainter. The authors further speculated that this might be due to a luminous X-ray 
emission component from the polar caps of radio-quiet pulsars missing the line-of-sight. 
Recently, Sokolova \& Rubtsov (2016) have also reported their attempt in searching the difference 
between radio-loud and radio-quiet populations. No significant differences in their ages and locations 
in the Galaxy have been found. On the other hand, there is a possible difference 
between their distributions of rotation period. 

The aforementioned studies have shown that the properties of radio-loud and radio-quiet
$\gamma-$ray pulsars can be intrinsically different. However, a thorough comparison of 
other characteristics of pulsars, such as magnetic field strength and 
spectral properties, remains unreported. This motivates us to perform a systematic search 
for the difference of the emission and physical properties between these two populations through 
a detailed statistical analysis.  

\section{Data Analysis}

All the data used in this work are collected from 2PC (Abdo et al. 2013) and the third \emph{Fermi} 
$\gamma-$ray point sources catalog (3FGL; Acero et al. 2015), which are summarized in 
Table~1 and Table~2. These parameters are chosen to characterize the pulsars in the following aspects:

\begin{enumerate}
\item
{\it Magnetic field strength and spin-down power} - Magnetic field strength is a crucial factor for the 
acceleration and emission processes in the magnetosphere (e.g. Cheng \& Zhang 1998). In this work, we compare the magnetic field 
of radio-loud and radio-quiet populations at the stellar surface $B_{\rm s}$ as well as at the light cylinder $B_{\rm LC}$. 
Their strength can be derived from the spin period $P$ and its first time derivative $\dot{P}$ 
as $B_{\rm s}=(2\pi)^{-1}(1.5I c^{3}P\dot{P})^{1/2}R_{\rm NS}^{-3}$ and 
$B_{\rm LC}=4\pi^{2}(1.5I\dot{P})^{1/2}(c^{3}P^{5})^{-1/2}$ respectively by assuming a dipolar field geometry,
where $I$, $R_{\rm NS}$ and $c$ are moment of inertia, stellar radius and the speed of light. We assume 
$I=10^{45}$~g~cm$^{2}$ and $R_{\rm NS}=10$~km throughout this work. 

We also compare the spin-down power $\dot{E}=4\pi^{2}I\dot{P}P^{-3}$ between these two populations. 
As the rotational energy of a neutron star provides the reservoir for the pulsar emission, 
both $\gamma-$ray and X-ray luminosities are found to be scaled with $\dot{E}$ 
(e.g. Abdo et al. 2013; Possenti et al. 2002). 

\item
{\it Emission and spectral properties} - The $\gamma-$ray spectra of pulsars are typically modeled by a form of 
a power-law with an exponential cut-off (PLE). The spectral shape of this model is characterized by two parameters, namely 
the photon index $\Gamma$ and the cut-off energy $E_{\rm cut}$. Such model is curved in comparison with a simple 
power-law (PL). The spectral curvature of the pulsars are quantified by the parameter \texttt{Curve$\_$Significance} in 3FGL, which 
are obtained by comparing the difference between the PLE and PL mode fittings (in unit of $\sigma$). 

Apart from comparing these spectral parameters between the radio-loud and radio-quiet 
population, we also compare their $\gamma-$ray-to-X-ray flux ratios $F_{\gamma}/F_{x}$. Although Marelli et al. (2015) 
have already pointed out the distributions of $F_{\gamma}/F_{x}$ are different between these two populations, an 
investigation of its possible correlation with other parameters such as $\dot{E}$ remains unreported. In this study, 
we will not consider the $\gamma-$ray luminosities of pulsars as they depends on the distances which have a large uncertainties, 
in particular for the radio-quiet population. 

\item
{\it Temporal properties} - The viewing geometry (i.e. the angle between the line-of-sight and the $\gamma-$ray emission regions)
can possibly be different between these two populations. This can possibly be reflected in their pulse profiles.
Different viewing geometry can lead to either be a large pulse width (FWHM) for the single peak cases\footnote{It is computed by the sum of 
\texttt{HWHM$\_$P1$\_$L} and \texttt{HWHM$\_$P1$\_$R} in 2PC.} or a large peak separation for the multiple peaks 
cases ($\Delta_{\gamma}$), depending on whether the line-of-sight cut through a single emission region or a multiple emission regions.
This motivates us to compare the combined distributions of FWHM and $\Delta_{\gamma}$ between 
these two populations. 

One of the radio-quiet pulsars PSR J2021+4026 has its $\gamma-$ray flux at energies
$>100$~MeV suddenly decreased by $\sim18\%$ near MJD~55850 (Allafort et al. 2013). This makes
it to be the first variable $\gamma-$ray pulsar has ever been observed. To investigate whether 
there is any difference between radio-quiet and radio loud populations in terms of the flux variability, 
we also compare their distributions of the parameter \texttt{Variability$\_$Index} in 3FGL. This parameter indicates the 
difference between the light curve of a source and its average flux level over the full time coverage in 3FGL (Acero et al. 2015). 
For a \texttt{Variability$\_$Index} larger than 72.44, the null hypothesis of a source being steady can be rejected at $99\%$ confidence 
level (Acero et al. 2015).
\end{enumerate}

\subsection{Anderson-Darling Test} 
The histograms and the cumulative distributions of the chosen parameters are shown in Figure 1 and 
Figure 2 respectively. For searching the possible differences between the radio-loud and radio-quiet populations, 
we apply the non-parametric two-sample Anderson-Darling (A-D) test (Anderson \& Darling 1952; Darling 1957; Pettitt 1976, Scholz \& Stephens 1987)
to their unbinned distributions (Figure 2). 

While Kolmogorov-Smirnov (K-S) test has been widely used to test whether two unbinned distributions are different, 
it is not sensitive to identify the difference locates at the edges of the distributions or when these two distributions
are crossed.\footnote{https://asaip.psu.edu/Articles/beware-the-kolmogorov-smirnov-test}. In view of this, we adopt 
A-D test in our analysis. Another advantage of A-D test over K-S test is the evidence that it is better capable of 
detecting small differences (Engmann \& Cousineau 2011). In this work, we perform the two-sample A-D test with the code 
implemented in \emph{scipy}.\footnote{https://www.scipy.org/} 
The results are summarized in Table~3. 

Among all the tested parameters, their distributions of \texttt{Curve$\_$Significance} are found to be the most incompatible
($p-$value$\sim0.0002$). This indicates the possible difference of their $\gamma-$ray spectral shape. 

For comparing their flux ratios $F_{\gamma}/F_{x}$, we omitted all the upper-limits in Tab.~1 and Tab.~2. A difference is found ($p-$value $\sim0.0005$), 
which is consistent with the conclusion reported by Marelli et al. (2015) based on comparing their binned histograms. 

Another interesting result comes from comparing the magnetic fields of these two populations. While we do not find any difference of 
surface field strength $B_{\rm S}$ between radio-loud and radio-quiet pulsars, 
the distributions of the magnetic field at the light cylinder $B_{\rm LC}$ are found to be different ($p-$value$\sim0.002$; see Fig. 1 \& 2). 

The statistical significances of the aforementioned differences are $\gtrsim3\sigma$. However, these results have not taken the uncertainties of the 
parameters into account. For the $F_{\gamma}/F_{x}$ reported by 2PC, their statistical uncertainties are rather large. The average percentage error is $\sim34\%$ and $\sim28\%$ 
in the radio-loud and radio-quiet populations respectively. Taking this into consideration, the difference of $F_{\gamma}/F_{x}$ 
between these two populations can be drastically reduced. Shifting their cumulative distributions within the tolerence of their statistical uncertainties, 
the difference can possibly be reconciled ($p-$value$\sim0.03$). 

For $B_{\rm LC}$, we estimate the uncertatinties by propagating the errors of $P$ and $\dot{P}$ reported in 2PC or the ATNF catalog (Manchester et al. 2005). 
The mean percentage errors of $B_{\rm LC}$ of radio-loud and radio-quiet populations are $\sim0.14\%$ and $\sim0.08\%$ respectively. In view of their small 
uncertainties, the statistical significance for the difference of $B_{\rm LC}$ between these two populations remains unaltered. 

In 3FGL, there is no error estimate for \texttt{Curve$\_$Significance}. However, the accuracy of this parameter depends on how well the
$\gamma-$ray spectra can be constrained so that one can discriminate whether PL or PLE models provide a better fit. This in turns depends on the 
photon statistics. Since radio-loud $\gamma-$ray pulsars can be more easily detected with the aid of their radio ephemeris, their detection significances 
are generally lower than that of radio-quiet $\gamma-$ray pulsars (see Tab. 1 and 2). 
Since it is more difficult to detect the faint pulsars at energies higher than the cut-off energy, this might lead to their apparently flatter spectra. 
In order to test the robustness for the difference of \texttt{Curve$\_$Significance} 
between these two populations, we alleviate this possible selection effect by re-running the A-D test on the pulsars detected at a level $>10\sigma$ 
(i.e. $TS>100$ in 3FGL). While all the radio-quiet pulsars satisfy this criteria, this reduces the sample size of the radio-loud pulsars to 29.  
In this case, the statistical significance for the difference of \texttt{Curve$\_$Significance} is reduced but remains marginally at a 
$\sim3\sigma$ level ($p-$value$\sim0.003$). 

We also considered if there is any selection effect can result in the observed difference in $B_{\rm LC}$. $B_{\rm LC}$ is a function of $P$ and $\dot{P}$. 
To investigate if the difference in $B_{\rm LC}$ is caused by the distributions of their rotational parameters, we have also applied the A-D test seperately on 
$P$ and $\dot{P}$. In the full sample, we have found a marginal difference of $P$ between this two populations ($p-$value$\sim0.006$). 
On the other hand, we do not find any difference in the distributions of $\dot{P}$ ($p-$value$\sim0.2$). 
However, we note that the difference in $P$ can possibly be a result of observational bias. For example, radio-loud pulsars can be found with 
their radio ephemeris. This might facilitate the detection of fast rotation. Attempting to alleviate such effect, Sokolova \& Rubtsov (2016) have 
constructed a bias-free sample by performing blind pulsar searches from all point sources in 3FGL using only LAT data. 
To estimate the impact of this possible selection effect in $P$ and $B_{LC}$, 
we re-run the A-D test on the pulsars (26 radio-quiet; 14 radio-loud) detected in the blind search by Sokolova \& Rubtsov (2016). 
We found that the statistical significance for the difference in $P$ is not undermined ($p-$value$\sim0.006$). 
For $B_{LC}$, we found the statistical significance for the difference between two populations may 
drop to the level of $\sim2.5\sigma$ ($p-$value$\sim0.01$).   

\subsection{Correlation \& Regression Analysis} 
In \S2.1, we have shown the possible differences between radio-loud and radio-quiet pulsars in terms of $F_{\gamma}/F_{x}$, 
\texttt{Curve$\_$Significance} and $B_{\rm LC}$. 
In order to test if there is any relation between the emission properties ($F_{\gamma}/F_{x}$,\texttt{Curve$\_$Significance}) 
and $B_{\rm LC}$ in each population. We proceed to perform the correlation analysis

From Figure~1, it is obvious that the distributions for most of these parameters do not resemble a Gaussian. In view of this, we adopt a 
non-parametric approach by computing the Spearman rank coefficients (Conover 1999; Siegel \& Castellan 1988). 2PC has also reported 
a possible correlation between the cut-off energy $E_{\rm cut}$ and $B_{\rm LC}$ for the radio-quiet $\gamma-$ray pulsars (Abdo et al. 2013). 
However, the authors have adopted 
a linear correlation analysis (i.e Pearson's $r$, Fisher 1944) which implicitly assumes $E_{\rm cut}$ and $B_{\rm LC}$ follow a bivariate 
Gaussian probability distribution. Such assumption is unlikely to be satisfied (cf. Fig.~1). Therefore, we have also run the non-parametric 
correlation analysis for $E_{\rm cut}-B_{\rm LC}$ to cross-check this possible relation. 
The results are summarized in Table~4.

For $F_{\gamma}/F_{x}$ and \texttt{Curve$\_$Significance}, we do not find any evidence for the correlation with $B_{\rm LC}$ in both radio-loud 
and radio-quiet populations. 
On the other hand, for the radio-quiet pulsars, $E_{\rm cut}$ is found to have a strong positive correlation with 
$B_{\rm LC}$ ($p-$value $\sim2\times10^{-6}$) 
However, this relation cannot be found in the radio-loud population ($p-$value $\sim0.1$). 

We further examine the phenomenological relation $E_{\rm cut}-B_{\rm LC}$ in the case of radio-quiet pulsars by assuming a linear model 
$E_{\rm cut}=a+b\log B_{\rm LC}$ in a regression analysis. The best-fit model is:

\begin{equation}
E_{\rm cut}=(-1.74\pm0.36)+(1.15\pm0.11)\log B_{\rm LC}~{\rm GeV},
\end{equation}

\noindent which is shown in Figure~4. The quoted uncertainties are 95$\%$ confidence intervals. We have also displayed the corresponding plot 
for the radio-loud pulsars for comparison. 

\section{Summary \& Discussions} 
We have performed a detailed statistical analysis to probe the physical nature of radio-loud and radio-quiet $\gamma-$ray pulsars. 
By comparing the cumulative frequency distributions of a set of selected parameters (see Figure 2), we have identified the possible differences 
between these two populations in several aspects (cf. Table 3). 
We found that the $\gamma-$ray spectral curvature of radio-quiet pulsars can 
be larger than that of radio-loud pulsars. 
While the surface magnetic field strength $B_{\rm s}$ has a similar distribution 
in both populations, their magnetic field strength at the light cylinder $B_{\rm LC}$ are found to be different. 
However, we need to point out that the significance can possibly be hampered by the effect of observational selection bias. 

In re-examining the distributions of nominal values of $F_{\gamma}/F_{x}$, 
we confirmed the difference between the radio-loud and radio-quiet pulsars as claimed by Marelli et al. (2015).  
However, with the large statistical uncertainties of $F_{\gamma}/F_{x}$ taking into account, it does not allow one to draw a
firm conclusion on their difference. 

While the possible differences identified in our analysis might be suffered from the selection effects and the statistical uncertainties, 
we note that such differences can be explained in the context of outer gap model by the geometric effect and the rotational period. 
In the following, we explain these properties 
qualitatively by assuming: (1) the $\gamma-$rays are originated from the outer gap, (2) the X-rays are originated from the polar cap due to backflow current 
heating, and (3) the open angle of the radio emission cone depends on $P^{-1/2}$ (e.g. Lyne \& Manchester 1988; Kijak \& Gil 1998, 2003). 

Since $B_{\rm LC}\sim B_{\rm s}P^{-3}$, the differences between radio-loud and radio-quiet populations
should stem from the rotational period $P$ (cf. Fig.~3). 
We noted that $P$ of radio-loud pulsars are generally smaller than radio-quiet pulsars. 
We first assume all pulsars have radio emission cones. Whether one is radio-loud or radio-quiet, it depends on whether the line-of-sight can meet 
the radio cone. 
From radio observations, it has been found that the radio cone size is related to the period of pulsars as $\sim P^{-\alpha}$ 
(e.g.,Narayan \& Vivekanand 1983; Lyne \& Manchester 1988; Biggs 1990; Gil, Kijak, \& Seiradakis 1993; Gil \& Han 1996; Kijak \& Gil 1998, 2003), 
where $\alpha$ is about 0.5. Therefore, shorter period pulsars will have wider radio cone and hence more favorable to be radio-loud. 
And hence the radio-quietness in the pulsar population might be a result of their narrower radio cones.  

Concerning the difference in $F_{\gamma}/F_{x}$, 
we consider a geometric effect together with assumption that the X-rays are coming from the regions near the polar cap 
(e.g. Arons 1981; Harding, Ozernoy, \&  Usov 1993; Cheng, Gil \& Zhang 1998; Cheng \& Zhang 1999).  
In this case its intensity $F_{x}^{\rm pc}$ should depend on the angle between the magnetic axis and the viewing angle $\theta$, 
namely $F_{x}^{\rm pc}\propto\cos\theta$. 
Based on the assumption that the line-of-sight of the radio-loud pulsars must be within the radio cone and that for radio-quiet pulsars is 
outside the radio cone, then radio loud pulsars should have smaller $\theta$ than those radio quiet pulsars. 
This implies the mean $F_{x}$ of radio-loud pulsars is larger than that of the radio-quiet pulsars. 
From observations and simulations (e.g. Takata, Wang and Cheng 2011), the difference in the $\gamma$-ray flux distributions between radio-loud and 
radio-quiet pulsars is not very large. On the other hand, $\cos\theta$ can vary from 0 to 1. 
Assuming $F_{\gamma}$ is similar for these two populations,
$F_{\gamma}/F_{x}$ of radio-quiet population should be larger than the radio-loud group. 

We note a special pulsar PSR J0537-6910 which is radio-quiet X-ray pulsar but without $\gamma-$ray emission detected 
(Marshall et al. 1998; Gotthelf et al. 1998). Its X-ray emission is likely to be non-thermal dominant (Gotthelf et al. 1998)
which presumably originated from the synchrotron emission of backflow current in the outer gap 
(cf. Cheng \& Zhang 1999). The non-detection of radio emission and the thermal
X-ray component imply that our line-of-sight is far from its polar cap region. As the beaming directions of the $\gamma-$rays and the non-thermal X-rays are
not necessary in the same direction (cf. Fig. 2 in Cheng \& Zhang 1999), our line-of-sight might miss the $\gamma-$ray emitting region as well.

To account for the difference of $\gamma-$ray spectral curvature, we speculate that 
inverse Compton (IC) process may play a role in high energy photon production. 
The most natural soft photons are radio. For the radio-loud pulsars, which generally have wider radio cones than their radio-quiet counterparts, 
part of radio emission with frequency $>100$ MHz may get into the outer gap 
and IC scatter with the primary electrons/positrons to the photons in GeV regime (cf. Ng et al. 2014). On the other hand, the probability of
radio photons in radio-quiet pulsars get into the gap is low. This could lead to a shortage of photons produced at higher energies through the 
aforementioned IC process. And this might result in more curved spectra of radio-quiet pulsars. 

$E_{\rm cut}$ of radio-quiet pulsars are found to be
strongly correlated with $B_{\rm LC}$. However, such association is absent in the radio-loud population. 
The aforementioned IC scenario can also provide a possible way to account for this phenonmena.
$E_{\rm cut}$ might be determined by IC scattering between the radio emission and the primary electrons/positrons in the outer gap. 
Such effect can be enhanced if the open angle of the radio cone is larger. And hence $E_{\rm cut}$ should be proportional to $1/P$ and 
this results in the positive correlation between $E_{\rm cut}$ and $B_{\rm LC}$. From the histograms (cf. Figure~1), 
we notice that the spread of $E_{\rm cut}$ is wider in the radio-loud population than that in the radio-quiet population. 
This might indicate that the factor of determining the cut-off energy is more complicated in the case of radio-loud pulsars. 

While the differences between the radio-quiet and radio-loud pulsars reported in this work are physically plausible by 
the outer gap model, a firm conclusion is limited by the current sample and various observational biases. 
With more $\gamma-$ray pulsars detected in the future, their properties suggested by our analysis can be re-examined. 

\begin{landscape}
\begin{table*}
\caption{The selected parameters of radio-loud $\gamma-$ray pulsars as described in the main text of Sec. 2.}
\centering
\resizebox{!}{7cm}{
\begin{tabular}{c c c c c c c c c c c c c c}
\hline
\hline
PSR & $P$ & $\dot{P}$ & $B_S$ & $B_{LC}$ & $\dot{E}$ & \texttt{Variability} & \texttt{Curve} & $E_{cut}$ & $\Gamma$ & $F_{\gamma}/F_X$ & $\Delta_{\gamma}$ & FWHM & $TS^{a}$ \\
 &  &  &  & & & \texttt{Index} & \texttt{Significance} &  & & & & & \\
 & (ms) & $(10^{-15}$ s/s) & $(10^{10}G)$ & (G)& ($10^{34}$ erg/s) & & & (GeV) & & & & & \\
\hline
J0205+6449 & 65.7 & 190 & 353.3 & 114617.6 & 2644 & 37.4 & 4.9 & $1.6\pm0.3$ & $1.8\pm0.1$ & $29.7\pm2.1$ & $0.503\pm0.004$ & $\cdot\cdot\cdot$ & 1019 \\
J0248+6021 & 217.1 & 55 & 345.6 & 3106.8 & 21.2 & 66.6 & 7.1 & $1.6\pm0.3$ & $1.8\pm0.1$ & $>57.4$ & $\cdot\cdot\cdot$ & 0.1968 & 578 \\
J0534+2200 & 33.6 & 420 & 375.7 & 911096.3 & 43606 & 621.9 & 15.8 & $4.2\pm0.2$ & $1.9\pm0.1$ & $0.296\pm0.007$ & $0.407\pm0.001$ & $\cdot\cdot\cdot$ & 102653 \\
J0631+1036 & 287.8 & 104 & 547.1 & 2111.4 & 17.3 & 42.5 & 8.0 & $6.0\pm1.0$ & $1.8\pm0.1$ & $>2070$ & $\cdot\cdot\cdot$ & 0.2216 & 621 \\
J0659+1414 & 384.9 & 55 & 460.1 & 742.3 & 3.8 & 45.3 & 7.3 & $0.4\pm0.2$ & $1.7\pm0.5$ & $61.8^{+6.3}_{-10.9}$ & $\cdot\cdot\cdot$ & 0.1596 & 419 \\
J0729-1448 & 251.7 & 114 & 535.7 & 3090.5 & 28.2 & 32.6 & 1.4 & $\cdot\cdot\cdot$ & $\cdot\cdot\cdot$ & $>318$ & $\cdot\cdot\cdot$ & 0.0423 & 54 \\
J0742-2822& 166.8 & 16.8 & 167.4 & 3318.6 & 14.3 & 58.3 & 4.1 & $1.6\pm0.8$ & $1.7\pm0.3$ & $>771$ &$\cdot\cdot\cdot$ & 0.0909 & 112 \\
J0835-4510 & 89.4 & 125 & 334.3 & 43042.7 & 690 & 20.0 & 54.0 & $3.0\pm0.1$ & $1.5\pm0.1$ & $1410\pm340$ & $0.433\pm0.001$ & $\cdot\cdot\cdot$ & 1659005 \\
J0908-4913 & 106.8 & 15.1 & 127.0 & 9590.7 & 49 & 47.3 & 1.9 & $0.5\pm0.2$ & $1.0\pm0.4$ & $>1130$ & $0.501\pm0.006$ & $\cdot\cdot\cdot$ & 315 \\
J0940-5428 & 87.6 & 32.8 & 169.5 & 23198.7 & 193 & $\cdot\cdot\cdot$ & $\cdot\cdot\cdot$ & $\cdot\cdot\cdot$ & $\cdot\cdot\cdot$ & $>314$ & $\cdot\cdot\cdot$ & 0.1631 & 14 \\
J1016-5857 & 107.4 & 80.6 & 294.2 & 21849.7 & 257 & 46.6 & 5.5 & $6.0\pm3.0$ & $1.8\pm0.2$ & $370^{+137}_{-343}$ & $0.423\pm0.004$ & $\cdot\cdot\cdot$ & 290 \\
J1019-5749 & 162.5 & 20.1 & 180.7 & 3874.8 & 18.4 & 63.7 & 3.1 & $\cdot\cdot\cdot$ & $\cdot\cdot\cdot$ & $>51.4$ & $\cdot\cdot\cdot$ & 0.0521 & 21 \\
J1028-5819 & 91.4 & 16.1 & 121.3 & 14616.2 & 83.3 & 71.1 & 21.3 & $4.6\pm0.5$ & $1.7\pm0.1$ & $5390\pm1660$ & $0.475\pm0.001$ & $\cdot\cdot\cdot$ & 5096 \\
J1048-5832 & 123.7 & 95.7 & 344.1 & 16723.2 & 200 & 56.6 & 18.1 & $3.0\pm0.3$ & $1.6\pm0.1$ & $4000^{+1490}_{-2800}$ & $0.426\pm0.001$ & $\cdot\cdot\cdot$ & 5389 \\
J1057-5226 & 197.1 & 5.8 & 106.9 & 1284.7 & 3 & 34.9 & 58.7 & $1.4\pm0.1$ & $1.0\pm0.1$ & $1950^{+40}_{-170}$ & $0.307\pm0.004$ & $\cdot\cdot\cdot$ & 27848 \\
J1105-6107 & 63.2 & 15.8 & 99.9 & 36418.6 & 248 & 56.1 & 1.8 & $1.3\pm0.6$ & $1.5\pm0.3$ & $>6130$ & $0.504\pm0.006$ & $\cdot\cdot\cdot$ & 309 \\
J1112-6103 & 65 & 31.5 & 143.1 & 47935.8 & 454 & 73.5 & 5.2 & $6.0\pm3.0$ & $1.6\pm0.3$ & $1070\pm560$ & $0.457\pm0.013$ & $\cdot\cdot\cdot$ & 58 \\
J1119-6127 & 408.7 & 4028 & 4057.4 & 5467.9 & 233 & 62.7 & 2.3 & $3.2\pm0.8$ & $1.8\pm0.1$ & $483\pm84$ & $0.204\pm0.02$ & $\cdot\cdot\cdot$ & 661 \\
J1124-5916 & 135.5 & 750 & 1008.1 & 37297.5 & 1190 & 36.0 & 8.5 & $2.1\pm0.4$ & $1.8\pm0.1$ & $63.1^{+9.5}_{-9.0}$ & $0.499\pm0.004$ & $\cdot\cdot\cdot$ & 1058 \\
J1357-6429 & 166.2 & 357 & 770.3 & 15436.3 & 307 & 54.6 & 2.9 & $0.9\pm0.5$ & $1.8\pm0.4$ & $809\pm324$ & $\cdot\cdot\cdot$ & 0.2637 & 187 \\
J1410-6132 & 50.1 & 31.8 & 126.2 & 92343.7 & 1000 & 35.4 & 2.8 & $\cdot\cdot\cdot$ & $\cdot\cdot\cdot$ & $>366$ & $0.458\pm0.037$ & $\cdot\cdot\cdot$ & 40 \\
J1420-6048 & 68.2 & 82.9 & 237.8 & 68961.1 & 1032 & 56.7 & 4.0 & $1.6\pm0.2$ & $1.5\pm0.1$ & $1060\pm480$ & $0.312\pm0.015$ & $\cdot\cdot\cdot$ & 1220 \\
J1509-5850 & 88.9 & 9.2 & 90.4 & 11842.1 & 51.5 & 52.7 & 10.6 & $4.6\pm0.9$ & $1.9\pm0.1$ & $2380^{+900}_{-830}$ & $0.264\pm0.013$ & $\cdot\cdot\cdot$ & 1152 \\
J1513-5908 & 151.5 & 1529 & 1522.0 & 40268.0 & 1735 & 60.2 & 0 & $\cdot\cdot\cdot$ & $\cdot\cdot\cdot$ & $0.612\pm0.284$ & $\cdot\cdot\cdot$ & 0.1912 & 98 \\
J1531-5610 & 84.2 & 13.8 & 107.8 & 16613.0 & 91.2 & $\cdot\cdot\cdot$ & $\cdot\cdot\cdot$ & $\cdot\cdot\cdot$ & $\cdot\cdot\cdot$ & $\cdot\cdot\cdot$ & $\cdot\cdot\cdot$ & 0.0607 & 2 \\
J1648-4611 & 165 & 23.7 & 197.7 & 4050.0 & 20.9 & 36.3 & 6.2 & $6.0\pm4.0$ & $1.6\pm0.3$ & $>2520$ & $0.298\pm0.082$ & $\cdot\cdot\cdot$ & 176 \\
J1702-4128 & 182.2 & 52.3 & 308.7 & 4695.3 & 34.2 & $\cdot\cdot\cdot$ & $\cdot\cdot\cdot$ & $0.8\pm0.5$ & $1.1\pm0.9$ & $3150^{+4500}_{-3150}$ & $\cdot\cdot\cdot$ & 0.2446 & 62 \\
J1709-4429 & 102.5 & 92.8 & 308.4 & 26348.3 & 340 & 54.1 & 28.5 & $4.2\pm0.1$ & $1.6\pm0.1$ & $3560^{+350}_{-890}$ & $0.244\pm0.002$ & $\cdot\cdot\cdot$ & 96893 \\
J1718-3825 & 74.7 & 13.2 & 99.3 & 21916.6 & 125 & 68.9 & 8.5 & $1.4\pm0.1$ & $1.5\pm0.1$ & $753^{+375}_{-622}$ & $\cdot\cdot\cdot$ & 0.1899 & 462 \\
J1730-3350 & 139.5 & 84.8 & 343.9 & 11656.0 & 123 & $\cdot\cdot\cdot$ & $\cdot\cdot\cdot$ & $1.2\pm0.3$ & $1.5\pm0.3$ & $>3280$ & $0.419\pm0.007$ & $\cdot\cdot\cdot$ & 100 \\
J1741-2054 & 413.7 & 17 & 265.2 & 344.6 & 0.9 & 48.8 & 25.1 & $0.9\pm0.1$ & $1.1\pm0.1$ & $187^{+13}_{-35}$ & $0.244\pm0.011$ & $\cdot\cdot\cdot$ & 3014 \\
J1747-2958 & 98.8 & 61.3 & 246.1 & 23476.1 & 251 & 60.1 & 11.8 & $1.9\pm0.1$ & $1.6\pm0.1$ & $43.3^{+19.2}_{-6.1}$ & $0.392\pm0.005$ & $\cdot\cdot\cdot$ & 1689 \\
J1801-2451 & 125 & 127 & 398.4 & 18767.9 & 257 & $\cdot\cdot\cdot$ & $\cdot\cdot\cdot$ & $3.0\pm2.0$ & $1.5\pm0.5$ & $75.3\pm45.9$ & $0.496\pm0.02$ & $\cdot\cdot\cdot$ & 58 \\
J1833-1034 & 61.9 & 202 & 353.6 & 137163 & 3364 & 56.0 & 3.5 & $0.9\pm0.2$ & $0.9\pm0.2$ & $8.89\pm1.15$ & $0.447\pm0.004$ & $\cdot\cdot\cdot$ & 258 \\
J1835-1106 & 165.9 & 20.6 & 184.9 & 3724.8 & 17.8 & $\cdot\cdot\cdot$ & $\cdot\cdot\cdot$ & $\cdot\cdot\cdot$ & $\cdot\cdot\cdot$ & $\cdot\cdot\cdot$ & $0.421\pm0.011$ & $\cdot\cdot\cdot$ & 30 \\
J1952+3252 & 39.5 & 5.8 & 47.9 & 71451.1 & 372 & 49.1 & 19.3 & $2.5\pm0.2$ & $1.5\pm0.1$ & $33.9\pm1.8$ & $0.478\pm0.003$ & $\cdot\cdot\cdot$ & 4469 \\
J2021+3651 & 103.7 & 95.6 & 314.9 & 25975.9 & 338 & 46.8 & 35.5 & $3.0\pm0.2$ & $1.7\pm0.1$ & $2300^{+260}_{-530}$ & $0.478\pm0.001$ & $\cdot\cdot\cdot$ & 17821 \\
J2030+3641 & 200.1 & 6.5 & 114.0 & 1309.6 & 3.2 & 31.1 & 12.1 & $1.5\pm0.4$ & $0.7\pm0.4$ & $>69.5$ & $0.309\pm0.014$ & $\cdot\cdot\cdot$ & 313 \\
J2032+4127 & 143.2 & 20.4 & 170.9 & 5354.8 & 27.3 & 38.3 & 15.5 & $3.2\pm0.5$ & $1.1\pm0.1$ & $5110^{+2630}_{-2950}$ & $0.516\pm0.001$ & $\cdot\cdot\cdot$ & 1383 \\
J2043+2740 & 96.1 & 1.2 & 34.0 & 3520.2 & 5.5 & 50.6 & 5.5 & $1.2\pm0.6$ & $1.4\pm0.4$ & $453^{+117}_{-255}$ & $0.432\pm0.01$ &$\cdot\cdot\cdot$ & 97 \\
J2229+6114 & 51.6 & 77.9 & 200.5 & 134255.6 & 2231 & 45.3 & 21.7 & $4.3\pm0.3$ & $1.8\pm0.1$ & $49.4^{+9.0}_{-5.7}$ & $0.299\pm0.008$ & $\cdot\cdot\cdot$ & 424 \\
J2240+5832 & 139.9 & 15.2 & 145.8 & 4899.7 & 21.9 & 52.8 & 5.3 & $3.0\pm2.0$ & $1.5\pm0.5$ & $>23.5$ & $0.476\pm0.014$ & $\cdot\cdot\cdot$ & 54 \\
\hline
\multicolumn{14}{l}{\footnotesize  (a) The test statistic ($TS$) values reported by 3FGL, which correspond to the detection significance $\sigma\simeq\sqrt{TS}$.}
 \end{tabular}

}
 \end{table*}
\end{landscape}

\begin{landscape}
\begin{table*}
\caption{ The selected parameters of radio-quiet $\gamma-$ray pulsars as described in the main text of Sec. 2.}
\centering
\resizebox{!}{6cm}{
\begin{tabular}{c c c c c c c c c c c c c c}
\hline
\hline
PSR & $P$ & $\dot{P}$ & $B_S$ & $B_{LC}$ & $\dot{E}$ & \texttt{Variability} & \texttt{Curve} & $E_{cut}$ & $\Gamma$ & $F_{\gamma}/F_X$ & $\Delta_{\gamma}$ & FWHM & $TS^{a}$ \\
 &  &  &  & & & \texttt{Index} & \texttt{Significance} &  & & & & & \\
 & (ms) & $(10^{-15}$ s/s) & $(10^{10}G)$ & (G)& ($10^{34}$ erg/s) & & & (GeV) & & & & & \\
 \hline
J0007+7303 & 315.9 & 357 & 1062.0 & 3099.2 & 44.8 & 46.2 & 22.7 & $4.7\pm0.2$ & $1.4\pm0.1$ & $4320\pm70$ & $0.216\pm0.005$ & $\cdot\cdot\cdot$ & 43388 \\
J0106+4855 & 83.2 & 0.43 & 18.9 & 3021.4 & 2.9 & 41.7 & 9.3 & $2.7\pm0.6$ & $1.2\pm0.2$ & $>229$ & $0.487\pm0.003$ & $\cdot\cdot\cdot$ & 544 \\
J0357+3205 & 444.1 & 13.1 & 241.2 & 253.4 & 0.6 & 47.8 & 22.7 & $0.8\pm0.1$ & $1.0\pm0.1$ & $1000^{+150}_{-100}$ & $\cdot\cdot\cdot$ & 0.2123 & 3468 \\
J0622+3749 & 333.2 & 25.4 & 290.9 & 723.5 & 2.7 & 54.0 & 9.7 & $0.6\pm0.1$ & $0.6\pm0.4$ & $>56.1$ & $0.457\pm0.034$ & $\cdot\cdot\cdot$ & 302 \\
J0633+0632 & 297.4 & 79.6 & 486.5 & 1701.7 & 11.9 & 59.4 & 17.3 & $2.7\pm0.3$ & $1.4\pm0.1$ & $1510\pm170$ & $0.476\pm0.003$ & $\cdot\cdot\cdot$ & 2448 \\
J0633+1746 & 237.1 & 11 & 161.5 & 1114.7 & 3.3 & 46.5 & 85.0 & $2.2\pm0.1$ & $1.2\pm0.1$ & $8520^{+160}_{-460}$ & $0.508\pm0.001$ & $\cdot\cdot\cdot$ & 906994 \\
J0734-1559 & 155.1 & 12.5 & 139.2 & 3433.3 & 13.2 & 31.9 & 10.2 & $3.2\pm0.9$ & $2.0\pm0.1$ & $>236$ & $ \cdot\cdot\cdot$ & 0.2627 & 916 \\
J1023-5746 & 111.5 & 382 & 652.6 & 43314.4 & 1089 & 53.7 & 15.3 & $2.5\pm0.4$ & $1.7\pm0.1$ & $2070^{+460}_{-1320}$ & $0.474\pm0.002$ & $\cdot\cdot\cdot$ & 2926 \\
J1044-5737 & 139 & 54.6 & 275.5 & 9437.3 & 80.2 & 60.0 & 15.7 & $2.8\pm0.3$ & $1.8\pm0.1$ & $1700^{+490}_{-1090}$ & $0.373\pm0.004$ & $\cdot\cdot\cdot$ & 3380 \\
J1135-6055 & 114.5 & 78.4 & 299.6 & 18362.5 & 206 & 46.4 & 9.0 & $2.4\pm0.5$ & $1.7\pm0.1$ & $1290^{+520}_{-1130}$ & $\cdot\cdot\cdot$ & 0.3138 & 498 \\
J1413-6205 & 109.7 & 27.4 & 173.4 & 12082.2 & 81.8 & 46.6 & 16.0 & $4.1\pm0.5$ & $1.5\pm0.1$ & $1120\pm310$ & $0.372\pm0.003$ & $\cdot\cdot\cdot$ & 1795 \\
J1418-6058 & 110.6 & 169 & 432.3 & 29399.7 & 494 & 65.3 & 16.1 & $5.5\pm0.5$ & $1.8\pm0.1$ & $8400\pm3420$ & $0.467\pm0.003$ & $\cdot\cdot\cdot$ & 3487 \\
J1429-5911 & 115.8 & 30.5 & 187.9 & 11134.4 & 77.4 & 48.3 & 14.6 & $2.2\pm0.3$ & $1.6\pm0.1$ & $>1100$ & $0.479\pm0.004$ & $\cdot\cdot\cdot$ & 822 \\
J1459-6053 & 103.2 & 25.3 & 161.6 & 13525.4 & 90.9 & 40.0 & 11.3 & $2.9\pm0.5$ & $2.0\pm0.1$ & $1520\pm420$ & $\cdot\cdot\cdot$ & 0.085 & 2046 \\
J1620-4927 & 171.9 & 10.5 & 134.3 & 2433.3 & 8.1 & 38.3 & 12.2 & $2.5\pm0.3$ & $1.3\pm0.1$ & $>2330$ & $0.231\pm0.03$ & $\cdot\cdot\cdot$ & 1407 \\
J1732-3131 & 196.5 & 28 & 234.6 & 2844.2 & 14.6 & 75.1 & 27.3 & $1.9\pm0.1$ & $1.0\pm0.1$ & $5260\pm1870$ & $0.419\pm0.002$ & $\cdot\cdot\cdot$ & 2821 \\
J1746-3239 & 199.5 & 6.6 & 114.7 & 1329.5 & 3.3 & 48.1 & 9.3 & $1.5\pm0.2$ & $1.4\pm0.1$ & $>416$ & $0.176\pm0.019$ & $\cdot\cdot\cdot$ & 654 \\
J1803-2149 & 106.3 & 19.5 & 144.0 & 11027.4 & 64.1 & 64.2 & 9.1 & $3.6\pm0.8$ & $1.6\pm0.1$ & $>2030$ & $0.394\pm0.009$ & $\cdot\cdot\cdot$ & 410 \\
J1809-2332 & 146.8 & 34.4 & 244.7 & 6535.1 & 43 & 34.4 & 30.2 & $ 3.4\pm0.2$ & $1.6\pm0.1$ & $3590\pm820$ & $0.358\pm0.002$ & $\cdot\cdot\cdot$ & 15781 \\
J1813-1246 & 48.1 & 17.6 & 92.0 & 76064.4 & 624 & 36.9 & 17.1 & $2.6\pm0.3$ & $1.9\pm0.1$ & $1840^{+330}_{-610}$ & $0.489\pm0.01$ & $\cdot\cdot\cdot$ & 4664 \\
J1826-1256 & 110.2 & 121 & 365.2 & 25103.1 & 358 & 51.9 & 24.0 & $2.2\pm0.2$ & $1.6\pm0.1$ & $3420\pm770$ & $0.48\pm0.001$ & $\cdot\cdot\cdot$ & 5160 \\
J1836+5925 & 173.3 & 1.5 & 51.0 & 901.2 & 1.1 & 43.1 & 71.8 & $2.0\pm0.1$ & $1.2\pm0.1$ & $19500^{+2300}_{-13400}$ & $0.537\pm0.006$ & $\cdot\cdot\cdot$ & 142427 \\
J1838-0537 & 145.7 & 465 & 823.1 & 24483.0 & 593 & 28.5 & 9.6 & $4.1\pm0.4$ & $1.6\pm0.1$ & $2130\pm230$ & $0.298\pm0.014$ & $\cdot\cdot\cdot$ & 1325 \\
J1846+0919 & 225.6 & 9.9 & 149.4 & 1197.5 & 3.4 & 58.9 & 10.7 & $2.2\pm0.5$ & $0.7\pm0.3$ & $>83.3$ & $0.244\pm0.022$ & $\cdot\cdot\cdot$ & 428 \\
J1907+0602 & 106.6 & 86.7 & 304.0 & 23089.0 & 282 & 70.1 & 18.1 & $2.9\pm0.3$ & $1.6\pm0.1$ & $4410\pm1050$ & $0.398\pm0.004$ & $\cdot\cdot\cdot$ & 3773 \\
J1954+2836 & 92.7 & 21.2 & 140.2 & 16190.4 & 105 & 53.3 & 13.6 & $3.3\pm0.4$ & $1.6\pm0.1$ & $>1370$ & $0.456\pm0.004$ & $\cdot\cdot\cdot$ & 1592 \\
J1957+5033 & 374.8 & 6.8 & 159.6 & 279.0 & 0.5 & 47.9 & 10.8 & $1.0\pm0.2$ & $1.3\pm0.2$ & $>810$ &  $\cdot\cdot\cdot$ & 0.2652 & 846 \\
J1958+2846 & 290.4 & 212 & 784.6 & 2947.6 & 34.2 & 51.7 & 15.4 & $2.0\pm0.3$ & $1.4\pm0.1$ & $667\pm325$ & $0.454\pm0.004$ & $\cdot\cdot\cdot$ & 1519 \\
J2021+4026 & 265.3 & 54.2 & 379.2 & 1868.3 & 11.4 & 157.7 & 58.8 & $2.6\pm0.1$ & $1.6\pm0.1$ & $64600\pm4000$ & $0.687\pm0.009$ & $\cdot\cdot\cdot$ & 53955 \\
J2028+3332 & 176.7 & 4.9 & 93.1 & 1551.7 & 3.5 & 51.2 & 12.3 & $1.9\pm0.3$  & $1.2\pm0.2$ & $>370$ & $0.451\pm0.003$ & $\cdot\cdot\cdot$ & 1058 \\
J2030+4415 & 227.1 & 6.5 & 121.5 & 954.3 & 2.2 & 36.4 & 11.7 &$1.7\pm0.3$ & $1.6\pm0.1$ & $>228$ & $0.505\pm0.007$ & $\cdot\cdot\cdot$ & 504 \\
J2055+2539 & 319.6 & 4.1 & 114.5 & 322.6 & 0.5 & 42.9 & 21.4 & $1.1\pm0.1$ & $1.0\pm0.1$ & $1240^{+350}_{-800}$ & $0.113\pm0.017$ & $\cdot\cdot\cdot$ & 2751 \\
J2111+4606 & 157.8 & 143 & 475.0 & 11122.1 & 144 & 46.2 & 8.0 & $5.0\pm1.0$ & $1.7\pm0.1$ & $>196$ & $0.337\pm0.011$ & $\cdot\cdot\cdot$ & 731 \\
J2139+4716 & 282.8 & 1.8 & 71.3 & 290.2 & 0.3 & 39.4 & 10.5 & $1.3\pm0.3$ & $1.3\pm0.2$ & $>73.1$ & $\cdot\cdot\cdot$ & 0.1434 & 369 \\
J2238+5903 & 162.7 & 97 & 397.3 & 8486.0 & 88.8 & 59.5 & 12.5 & $2.1\pm0.3$ & $1.6\pm0.1$ & $>143$ & $0.502\pm0.002$ & $\cdot\cdot\cdot$ & 1165 \\
\hline
\multicolumn{14}{l}{\footnotesize (a) The test statistic ($TS$) values reported by 3FGL, which correspond to the detection significance $\sigma\simeq\sqrt{TS}$.}
 \end{tabular}
}
 \end{table*}
\end{landscape}

\begin{table*}
\caption{Summary of the results of A-D tests.}
\begin{center}
\begin{tabular}{c c c}
\hline
\hline
  & Null hypothesis probability$^{a}$ \\
\hline
$P$  & 0.006 \\
$\dot{P}$  & 0.2 \\
$B_S$  & 0.8 \\
$B_{LC}$  & 0.002 \\
$F_\gamma/F_X$  & 0.0005 \\
$\dot{E}$ & 0.003 \\
$\texttt{Curve$\_$Significance}$  & 0.0002 \\
$\texttt{Variability$\_$Index}$  & 0.7 \\
$\Gamma$  & 0.3 \\
$E_{\rm cut}$  & 0.1 \\
$\Delta_\gamma \cup FWHM$  & 0.4 \\
\hline
\\
\end{tabular}
\end{center}
\footnotesize{$(a)$ The probability of obtaining the two-sample A-D statistic larger or equal to the observed value under the 
null hypothesis that the distributions of the corresponding properties of radio-loud and radio-quiet $\gamma-$ray pulsars 
are the same.}
 \end{table*}

\begin{table*}
\caption{Summary of the results of correlation analysis}
\begin{center}
\begin{tabular}{c|cc}
\hline
\hline
Relation  &  Spearman Rank  & Probability$^{a}$ \\
\hline
\multicolumn{3}{c}{Radio-loud pulsar population}\\
\hline
$F_\gamma/F_X$ vs. $B_{\rm LC}$  & -0.5 &  0.01 \\
\texttt{Curve$\_$Significance} vs. $B_{\rm LC}$  & -0.03  & 0.8 \\
$E_{\rm cut}$ vs. $B_{\rm LC}$  & 0.3  & 0.1 \\
\hline
\multicolumn{3}{c}{Radio-quiet pulsar population}\\
\hline
$F_\gamma/F_X$ vs. $B_{\rm LC}$  & 0.04 & 0.9 \\
\texttt{Curve$\_$Significance} vs. $B_{\rm LC}$  & -0.06 & 0.7 \\
$E_{\rm cut}$ vs. $B_{\rm LC}$  & 0.7  & $2\times10^{-6}$ \\
\hline
\hline
 \end{tabular}
\end{center}
\footnotesize{$(a)$ The probability of obtaining the Spearman rank at least as extreme as the observed value under the 
hypothesis that there is no correlation between the tested pair of parameters. No assumption on the distributions of the parameters is 
required.}\\
 \end{table*}

\begin{figure*}[t]
\centerline{\psfig{figure=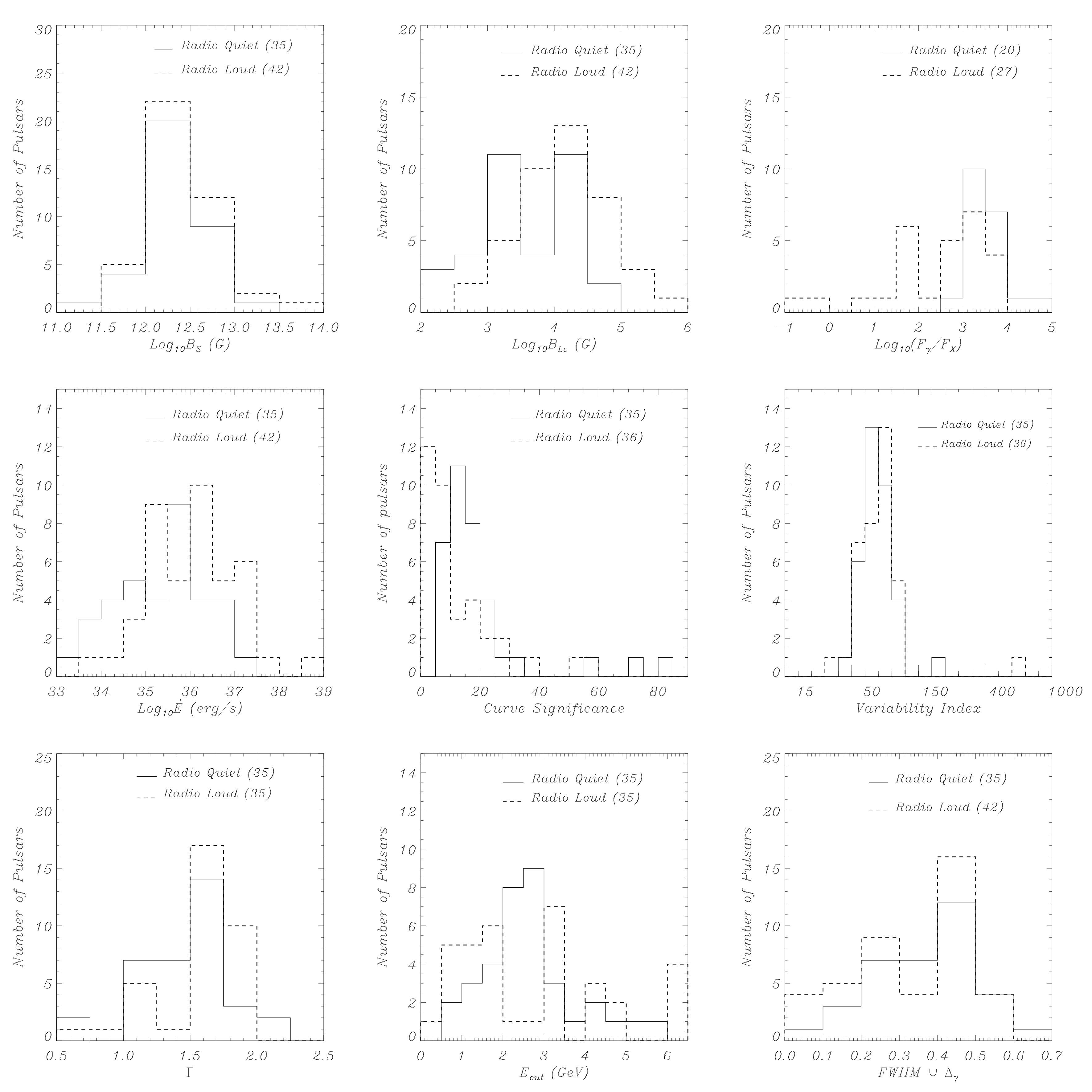,width=17cm,clip=}}
\caption[]{Histograms of the selected parameters for radio-loud ({\it dashed lines)} and radio-quiet ({\it solid lines}) $\gamma-$ray pulsars. 
The numbers in the parentheses are the sample sizes for the corresponding distributions.} 
\label{histo}
\end{figure*}

\begin{figure*}[t]
\centerline{\psfig{figure=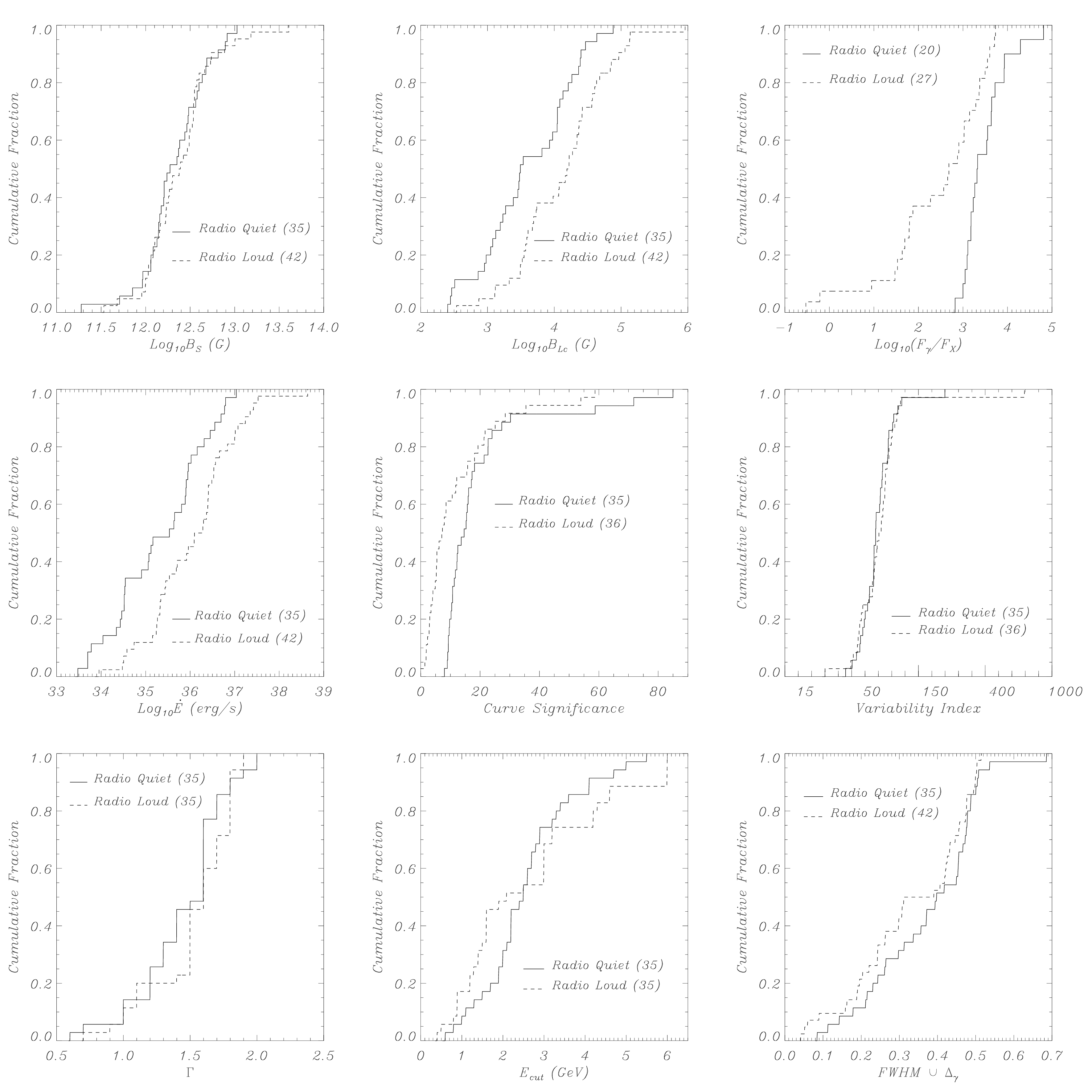,width=17cm,clip=}}
\caption[]{Cumulative frequency distributions of the selected parameters for radio-loud 
({\it dashed lines)} and radio-quiet ({\it solid lines}) $\gamma-$ray pulsars.  
The numbers in the parentheses are the sample sizes for the corresponding distributions.} 
\label{cumul}
\end{figure*}

\begin{figure*}[t]
\centerline{\psfig{figure=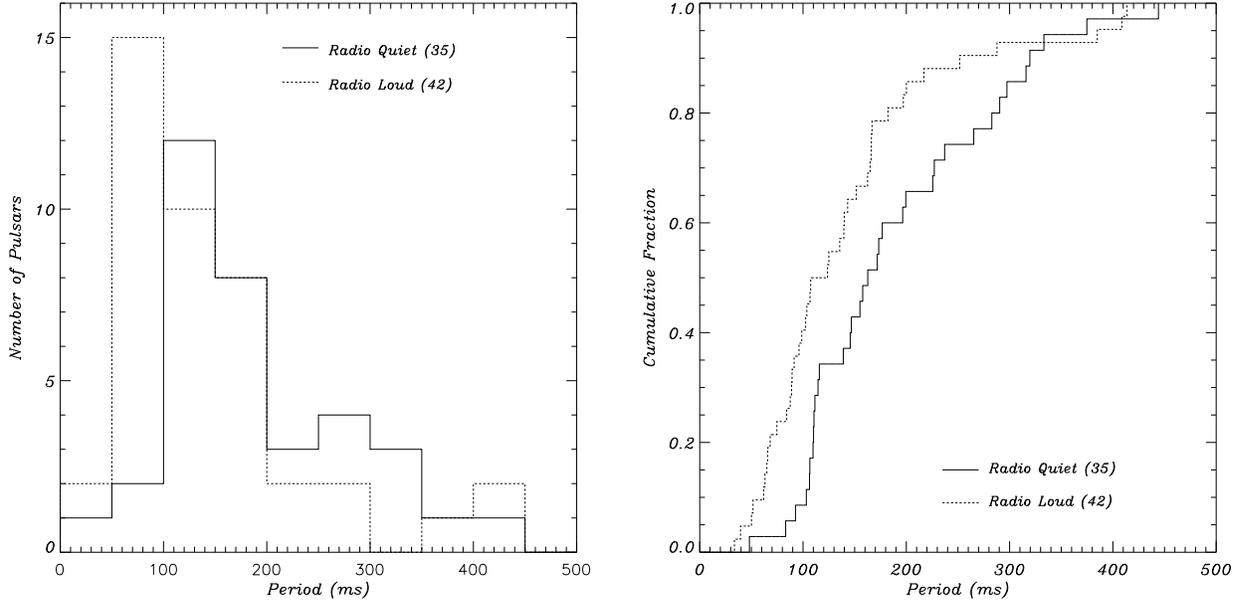,width=17cm,clip=}}
\caption[]{Comparing the rotational period distributions from the radio-loud and radio-quiet $\gamma-$ray pulsars in histograms ({\it left panel}) and 
cumulative distributions ({\it right panel}).} 
\label{Period}
\end{figure*}

\begin{figure*}[t]
\centerline{\psfig{figure=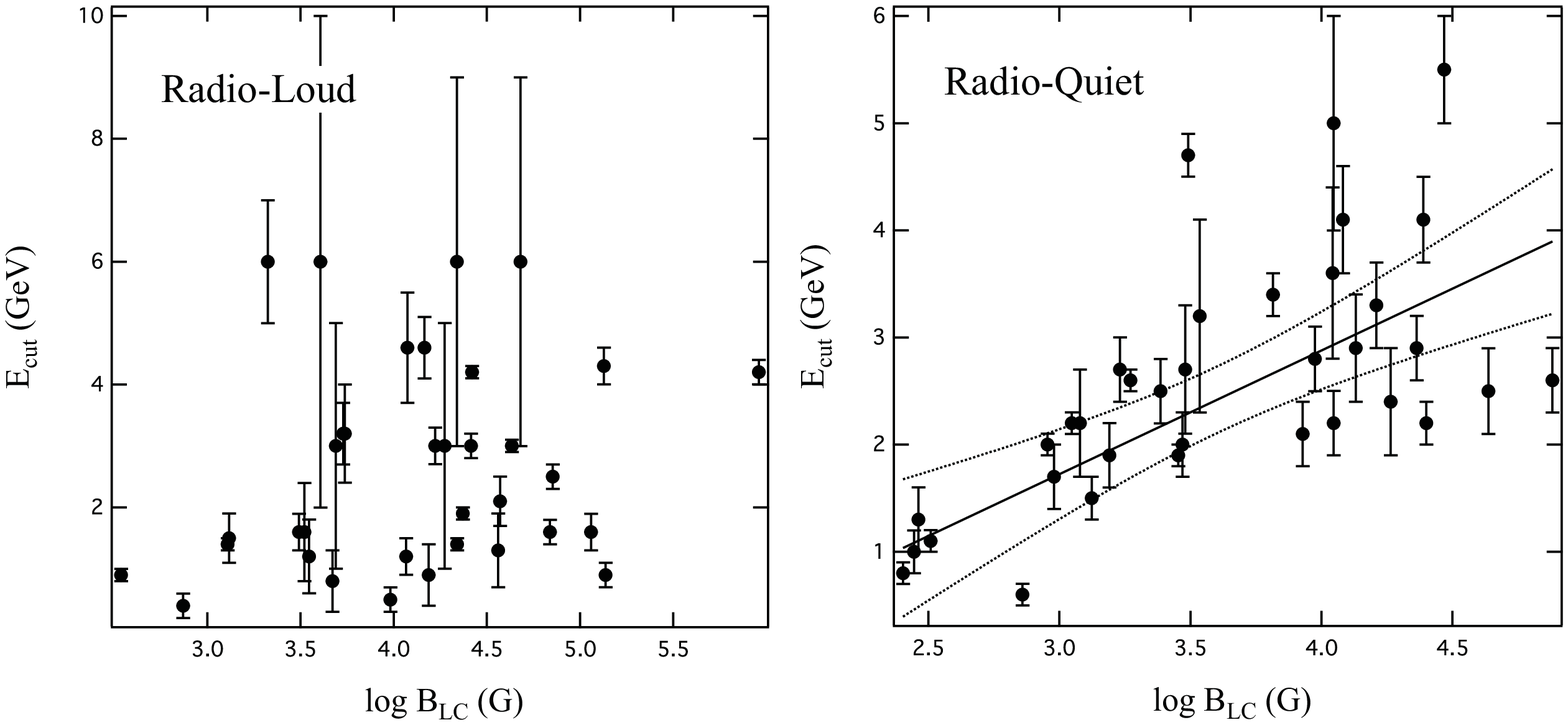,width=17cm,clip=}}
\caption[]{The cut-off energies $E_{\rm cut}$ vs. $B_{\rm LC}$ in radio-loud ({\it left panel}) and radio-quiet ({\it right panel}) 
$\gamma-$ray pulsar populations. The solid line in the right panel represent the best-fit from the regression analysis. 
The dotted lines represent the upper and lower $95\%$ confidence bands.} 
\label{Ecut}
\end{figure*}

\acknowledgments{\small
CYH is supported by the research fund of Chungnam National University. 
JL is is supported by BK21 plus program and 2016R1A5A1013277
JT is supported by the NSFC grants of China under 11573010
KSC are supported by a 2014 GRF grant of Hong Kong Government under HKU 17300814P. 
}



\begin{thebibliography}{}
\bibitem{} Abdo, A.~A., Ackermann, M., Ajello, M., et al. 2009a, Science, 325, 840
\bibitem{} Abdo, A.~A., Ackermann, M., Ajello, M., et al. 2009b, Science, 325, 848
\bibitem{} Abdo, A. A., Ajello, M., Allafort, A., et al. 2013, ApJS, 208, 17
\bibitem{} Acero, F., Ackermann, M., Ajello, M. et al. 2015, ApJS, 218, 23
\bibitem{} Anderson, T.W., \& Darling, D.A. 1952, Annals of Mathematical Statistics, 23 193
\bibitem{} Arons, J. 1981, ApJ, 248, 1099
\bibitem{} Arons, J. 1996, A\&AS, 120, C49
\bibitem{} Allafort, A., et al. 2013, ApJ, 777, L2
\bibitem{} Biggs, J. D. 1990, MNRAS, 245, 514
\bibitem{} Bignami, G. F., \& Caraveo, P. A. 1996, ARA\&A, 1996, 34, 331
\bibitem{} Cheng, K.~S., Gil, J. A., \& Zhang, L. 1998, ApJ, 493, L35
\bibitem{} Cheng, K.~S., \& Zhang, L. 1999, \apj, 515, 337
\bibitem{} Cheng, K.~S., \& Zhang, L. 1998, \apj, 498, 327
\bibitem{} Conover, W.J. 1999, Practical Nonparametric Statistics, 3rd Ed. Wiley
\bibitem{} Darling, D.A. 1957, Annals of Mathematical Statistics, 28, 823
\bibitem{} Engmann, S., \& Consineau, D. 2011, Journal of Applied Quantitative Methods, 6, 1
\bibitem{} Fisher, R.A. 1944, Statistical Methods for Research Workers, Oliver \& Boyd
\bibitem{} Gil, J. A., \& Han, J. L. 1996, ApJ, 458, 265
\bibitem{} Gil, J. A., Kijak, J., \& Seiradakis, J. H. 1993, A\&A, 272, 268
\bibitem{} Gotthelf, E. V., Zhang, W., Marshall, F. E., Middleditch, J., \& Wang, Q. D. 1998, MmSAI, 69, 825
\bibitem{} Harding, A. K., Ozernoy, L. M., \& Usov, V. V. 1993, MNRAS, 265, 921
\bibitem{} Kerr, M. 2011, \apj, 732, 38
\bibitem{} Kijak, J., \& Gil, J. 1998, MNRAS, 299, 855
\bibitem{} Kijak, J., \& Gil, J. 2003, A\&A, 397, 969
\bibitem{} Lyne, A. G., \& Manchester, R. N. 1988, MNRAS, 234, 477
\bibitem{} Marshall, F. E., Gotthelf, E. V., Zhang, W., Middleditch, J., \& Wang, Q. D. 1998, ApJ, 499, L179
\bibitem{} Marelli, M., et al. 2015, ApJ, 802, 78
\bibitem{} Marelli, M. 2012, PhD Thesis, University of Insubria
\bibitem{} Marelli, M., De Luca, A., \& Caraveo, P. A. 2011, ApJ, 733, 82
\bibitem{} Narayan, R., \& Vivekanand, M. 1983, A\&A, 122, 45
\bibitem{} Ng, C.-Y., et al. 2014, ApJ, 787, 167 
\bibitem{} Pettitt, A.N. 1976, Biometrika, 63, 161
\bibitem{} Possenti A., Cerutti R., Colpi M., Mereghetti S., 2002, A\&A, 387, 993
\bibitem{} Scholz, F.W., \& Stephen, M.A. 1987, Journal of the American Statistical Association, 82(339), 918
\bibitem{} Siegel, S., \& Castellan, N.J. 1988, Nonparametric Statistics for Behavioural Sciences, McGraw-Hill
\bibitem{} Sokolova, E. V., \& Rubtsov, G. I. 2016, arXiv:1601.00330
\bibitem{} Takata, J., Chang, H., \& Shibata, S. 2008, \mnras, 386, 748
\bibitem{} Takata, J., Shibata, S., Hirotani, K., \& Chang, H.-K. 2006, \mnras, 366, 1310
\bibitem{} Takata, J., Wang, Y., \& Cheng, K. S. 2011, ApJ, 726, 44
\bibitem{} Thompson, D. J. 2008, Rep. Prog. Phys., 71, 116901
\end{thebibliography}
\end{document}